\documentclass{bioinfo}
\copyrightyear{2015} \pubyear{2015}

\access{Advance Access Publication Date: Day Month Year}
\appnotes{Manuscript Category}

\begin{document}
\firstpage{1}

\subtitle{Genome analysis}

\title[Estimation of branching models for transposable elements dynamics]{Simulation based estimation of branching models for LTR retrotransposons}

\author[Moulin \textit{et~al}.]{Serge Moulin\,$^{\text{\sfb 1}*}$, Nicolas Seux\,$^{\text{\sfb 2}}$, St\'ephane Chr\'etien\,$^{\text{\sfb 3}}$, Christophe Guyeux\,$^{\text{\sfb 1}}$, and Emmanuelle Lerat\,$^{\text{\sfb 4,}}$}

\address{$^{\text{\sf 1}}$D\'epartement d'Informatique des Syst\`emes Complexes, UMR 6174 CNRS, FEMTO-ST Institute, 15 Bis Avenue des Montboucons, 25030 Besan\c con, France \\
$^{\text{\sf 2}}$Laboratoire de Math\'ematiques, Universit\'e de Franche-Comt\'e, UMR 6623 CNRS, 16 route de Gray, 25030 Besan\c con, France
$^{\text{\sf 3}}$National Physical Laboratory, Hampton Road, Teddington, United Kingdom and\\
$^{\text{\sf 4}}$Laboratoire Biometrie et Biologie Evolutive, Universite Claude Bernard - Lyon 1, UMR 5558 CNRS, 43 boulevard du 11 novembre 1918, 69622 Villeurbanne, France}

\corresp{}

\history{Received on XXXXX; revised on XXXXX; accepted on XXXXX}

\editor{Associate Editor: XXXXXXX}

\abstract{\textbf{Motivation:} LTR retrotransposons are mobile elements that are able, like retroviruses, to copy and move inside eukaryotic genomes. In the present work, we propose a branching model for studying the propagation of LTR retrotransposons in these genomes. This model allows to take into account both positions and degradations of LTR retrotransposons copies. In our model, the duplication rate is also allowed to vary with the degradation level.\\
\textbf{Results:} Various functions have been implemented in order to simulate their spread and visualization tools are proposed.
Based on these simulation tools, we show that an accurate estimation of the parameters of this propagation model can be performed.  
We applied this method to the study of the spread of the transposable elements ROO, GYPSY, and DM412 on a chromosome of \textit{Drosophila melanogaster}. \\
\textbf{Availability:} Our proposal has been implemented using Python software. Source code is freely available on the web at https://github.com/SergeMOULIN/retrotransposons-spread. \\
\textbf{Contact:} \href{serge.moulin@univ-fcomte.fr}{serge.moulin@univ-fcomte.fr}}

\maketitle

\section{Introduction}
\label{sec:Introduction}
A transposable element (TE) is a DNA sequence able to move from one location to another inside a genome. These sequences, discovered during the 50's by Barbara McClintock \citep{barbara1987discovery} exist in almost all living organisms and are the source of a huge number of mutations. They are considered as a major cause of genetic disease in human \citep{belancio2008mammalian} or in drosophila where they are responsible for more than 80\% of the spontaneous mutations \citep{green1988mobile}. DNA sequences derived from these TEs can represent a large part of a genome. For example they represent about 45\% of the human genome \citep{lander2001initial} and over 70\% of the corn genome \citep{sanmiguel1998evidence}. Hopefully, most of these sequences correspond to fragments or ``dead'' elements that have lost their ability to move in the genome due to several lethal mutations or are controlled especially via epigenetic mechanisms.

TEs have two possible ways to move in a genome, according to their type \citep{wicker2007unified}. The first class of mobile elements are cut from their original place to move to another one, and are called ``DNA transposons'' or ``Class II transposable elements''. The other class of mobile elements, called ``retrotransposons'' or ``Class I transposable elements'', use an RNA intermediate to duplicate themselves, the new copy being inserted into another location of the genome. Two orders are identified among the retrotransposons according to the presence or absence of Long Terminal Repeat (LTR) sequences at their extremities. The LTR retrotransposons are similar in structure to retroviruses such as HIV. In both classes, TEs can be classified as either ``autonomous'', if they code the enzymes that will allow them to move, or ``non autonomous'' if they use the enzymes produced by other elements. In an assembled genome, the various sequences corresponding to TE insertions can be found using different bioinformatic approaches (see \citep{lerat2011comparative} for a review), which allow to determine the exact number and positions of each TE insertion. In this article, we focused on the important problem of inferring the history of the spreading of LTR retrotransposons. For this purpose, we modeled the evolution using a branching process where each element (\textit{i.e.}, a copy of some TE) can randomly evolve via duplication or mutation.

Instances of branching processes have been proposed in the literature as models for the propagation of TEs. To the best of our knowledge, the first model of that kind has been designed by Kaplan \textit{et al.} in 1985 \citep{kaplan1985evolution}. In this paper, the authors proposed a model where TEs can be either of wild type (\textit{i.e.}, non mutated) or of mutant type. At each generation, wild copies can mutate or disappear, whereas mutant ones can only disappear, and the number of new copies created by transposition at each generation is assumed to follow a Poisson distribution. This number is supposed to decrease with the number of surviving copies, in order to account for the reducing number of unoccupied sites where transposition could take place. This number of new copies is also assumed to be a decreasing function of the proportion of mutants. Finally, the proportion of mutant type among newly created copies increases with the proportion of mutant type in existing copies. 
The temporal behavior of the number of copies of each type has been studied as well as the extinction time of the TE families.

In 1988, Michael E. Moody \citep{moody1988branching} has used a branching model to describe the propagation of TEs in a haploid population. The variable studied was the number of individuals possessing $i$ copies of a given TE (denoted by $Z^{(i)}_t$ for the $t^{th}$ generation). This number of copies was itself limited by the number $m$ of viable sites where TE could settle down. This model incorporated a copy-dependent selection. That is to say, the number of identical offspring produced by an individual is supposed to follow a law $Y^{(i)}$  dependent on $i$. In addition, some probabilities were defined, $p_i$, $q_i$, and 1 - $p_i$ - $q_i$ for an individual to: gain a copy of the TE, lose one element, or remain stable ($q_m = 0$). Another probability was defined $\beta$ for an individual devoid of the element to acquire a single copy. Finally, the asymptotic stability of $Z_t = (Z^{(0)}, ..., Z^{(max)})$ was studied based on these factors. This study shows that the TE load can maintain itself at intermediate level even in case of a neutral selection. In particular, models with copy-dependent transposition rate can provide a stationary distribution when the deletion rate is between the extremes of transposition rate.

Sawyer \textit{et al.} \citep{sawyer1987distribution} produced a model very close to Moody's in order to study the distribution and abundance of insertion sequences, which are DNA transposons, among natural isolates of \textit{Escherichia coli}. The main difference is that deletion is not taken into account. The reason for this choice is that the data available seemed to indicate that deletion occurs at a substantially smaller rate than transposition \citep{egner1981excision, foster1981three}.

More recently, interesting models have been proposed that take into account the location of TEs. For instance, Drakos and Wahl \citep{drakos2015extinction} suggested a model of mobile promoter evolution where the probabilities for promoters to duplicate inside or outside their region were potentially different. 

In the present work, the objective is to combine a location dependent model with the fact that LTR retrotransposons can face degradation, which may decrease their duplication rate. 
This model assumes that each copy can be either duplicated or mutated at any time. The time before duplication is supposed to follow an exponential distribution, as well as the time before mutation, the impact of mutation, and the distance covered by a duplicated copy before insertion. This model also takes into account the position of the root (\textit{i.e.} the first copy), and the impact of mutations faced by a TE on its duplication speed.
Our main goal is to estimate the parameters for this model. For this purpose, we propose a simulation based procedure. 
This method requires to define a distance between the results of the simulations and the observed genome, which is based on the Hungarian method \citep{kuhn1955hungarian, munkres1957algorithms}. This method has been applied to the spreading of the LTR retrotransposons ROO, DM412, and GYPSY on the chromosome 3L of \textit{Drosophila melanogaster}. The parameters associated to each TE are computed and a branching tree is proposed in each case. Our results show that, according to our model and method, the roots of ROO, DM412, and GYSPSY on the chromosome 3L could correspond respectively to the annotated copies FBti0060418, FBti0020009, and FBti0020033.

\section{System and methods}
\subsection{The branching model}
\label{sec:branching model}
\subsubsection{The branching tree}
\label{sec:branching tree}

\begin{figure}[!tpb]
\centerline{\includegraphics[scale=0.45]{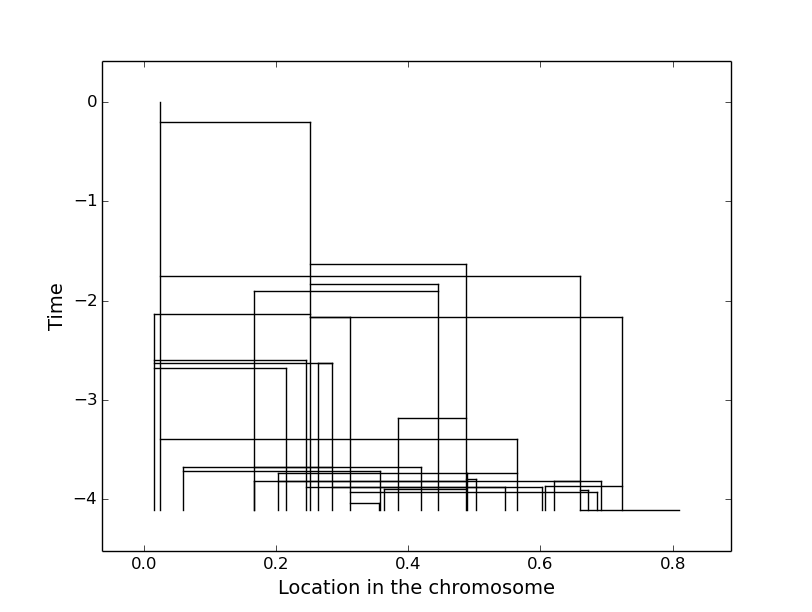}}
\caption{ROO spread}
\label{fig:ROO_spread}
\end{figure}

An example of branching tree is shown in Figure~\ref{fig:ROO_spread}.

This branching tree represents the spread of the LTR retrotransposons ``ROO''
between times 0 and 4.105.
At time 0 there is only one copy, called the ``root'' in this article. At time 0.205, the root duplicates itself to give birth to its first ``child''. Finally, the genome is observed at time $T_{obs}$ = 4.105 with 32 copies of ``ROO'' inserted in it.
The state of the tree at time $T_{obs}$ = 4.105 is named ``final state'' of this tree. The working principle of our estimation method is to simulate trees in order to determine in which conditions final states of simulated trees match well with the observed genome. 
The branching tree represents only duplications, but copies are also subject to mutations as explained in Section~\ref{sec:general model}.
To compare the final state of a simulated tree with the observed genome, the copy locations in the genome have been considered as well as their mutations, as detailed in Section~\ref{sec:Distance}.

\subsubsection{The general model}
\label{sec:general model}
The branching model is constructed as follows.
\begin{enumerate}

\item The spread starts with only one copy, called the root, at time zero in a location $X_0$ to determine. 

\item Each copy can be either duplicated or mutated at any time.

\item The number of copies increases due to duplications. When a new copy is created, it receives an index equal to the number of existing copies 
at the time of its birth, including itself. In the remainder of this article, let $T_i$ be the birth date of the $i^{th}$ copy and let $\tau_{i,k}$ be the time when the $i^{th}$ copy faces its $k^{th}$ mutation.

\item The time intervals $\tau_{i,k+1}-\tau_{i,k}$ between two mutations is supposed to be independent and identically distributed (i.i.d.) with exponential distribution $\mathcal E(\frac{1}{\mu})$, where $\mu$ (\textit{i.e.}, average time between two mutations) must be determined.

\item The proportion of nucleotides affected by a mutation is supposed to follow a distribution min$(1, \mathcal E(\frac{1}{\beta}))$, where $\beta$ must be determined.

\item Each copy is also associated to its Needleman-Wunsch \citep{needleman1970general} distance to the original state of the root. This distance, also named ``state of deterioration'' in the remainder of this article, is denoted by $D_i(t)$ for the $i^{th}$ copy at time $t$. This distance increases as a function of time, due to mutation effects. Finally, $D_i(T_{obs})$ is the state of deterioration at the end of the spread.

\item At time $t$, for the $i^{th}$ copy, conditionally on $D_i(t)$, we assume that the time before the next duplication follows a distribution $\mathcal E(\frac{1}{1 + p \times D_i(t)})$, where $p > 0$ is a parameter to determine. 
In other words, the time before the next duplication is longer when the copy is far from the original state of the root in terms of Needleman-Wunsch distance. 

\item Moreover, each copy is also associated to its position in the genome. This position is denoted by $X_i$ for the $i^{th}$ copy. 
This position is constant with time. We assume that each child $j$ of a copy $i$ satisfies $X_{j} = X_{i} + \chi_{i,j}$, where $\chi_{i,j}$  follows a distribution $\mathcal U\left\{-1,1\right\} \times \mathcal E (\frac{1}{L})$, in which $\mathcal U$ represents the uniform law (\textit{i.e.}, the probability to choose -1 or 1 is the same) and $L$ is a parameter to determine.
\end{enumerate}

Our goal is thus to estimate the parameters of this model, \textit{i.e.}, $ X_0, \mu, \beta, p$, and $L$. 
Note that the duplication speed of the non-mutated root is set to 1 and it does not need to be determined. Indeed, this duplication speed is redundant with $\mu$ and $p$.

\subsection{The estimation method}
\label{sec:method}
As explained in Section~\ref{sec:branching tree}, the working principle of our estimation method is to simulate trees in order to determine in which conditions the final states of simulated trees match well with the observed genome.

Trees are simulated according to the model defined in Section~\ref{sec:general model}. The stopping criterion of these simulated trees depends on the number of copies in the observed genome. Actually the simulation was constrained to stop randomly between the birth date of the $n^{th}$ and the birth date of the $n+1^{th}$ copy, where $n$ is the number of copies in the observed genome, see Section~\ref{sec:Stop condition} for further details. 

In what follows, Section~\ref{sec:Distance} defines the distance between the final state of a simulated tree and the observed genome, and Section~\ref{sec:Minimisation} presents how this distance has been minimized. 

\subsubsection{Distance between trees}
\label{sec:Distance}

The first step is to define a distance between the final state of a simulated tree and the observed genome, that is to say, a distance between final states of two trees. 
For this purpose, a distance between two copies has been designed as follows.
\begin{center}$D(R_i, \widetilde{R}_j) = |X_i - \widetilde{X}_j| + |D_i(T_{obs}) - \widetilde{D}_j(T_{obs})|$ \end{center} where $R_i$ is the $i^{th}$ copy of the first tree and $\widetilde{R}_j$ the $j^{th}$ copy of the second one. 
$|X_i - \widetilde{X}_j|$ represents here the geographical distance between $R_i$ and $\widetilde{R}_j$ while $D_i(T_{obs})$ is for the Needleman-Wunsch distance between $R_i$ and the state of the root at the origin. 
Then, the distance between final states of two trees has been defined as the best possible adjustment between copies, using Kuhn-Munkres algorithm (also named Hungarian method) \citep{kuhn1955hungarian, munkres1957algorithms}.

To illustrate this distance, let us consider the following example: at its final state, the first tree has 3 copies of a given TE. Their respective locations in the genome are $X_1$ = 0.58,  $X_2$ = 0.03, and $X_3$ = 0.97 while their respective Needleman-Wunsch distances from the original state of the root are $D_1(1)$ = 0.04, $D_2(1)$ = 0.13, and $D_3(1)$ =  0.14. The second tree has also 3 copies. Their respective locations in the genome are $\widetilde{X}_1$ = 0.55, $\widetilde{X}_2$ = 0.90, and $\widetilde{X}_3$ = 0.96 while their respective Needleman-Wunsch distances from the original state of the root are $D_1(1)$ = 0.07, $D_2(1)$ = 0.06, and $D_3(1)$ = 0.08.

The matrix of distances between copies of the two trees can thus be constructed as in Table~\ref{tab:Matrix}.

\begin{table}[!t]
\processtable{The matrix of distance between copies\label{tab:Matrix}}
{\begin{tabular}{@{}l|lll|@{}} \toprule
&$X_1$ = 0.58     &$X_2$ = 0.03     &$X_3$  = 0.97  \\
&$D_1(T_{obs})$ = 0.04  &$D_2(T_{obs})$ = 0.13  &$D_3(T_{obs})$ =  0.14 \\
\midrule
$\widetilde{X}_1$ =0.55  $\widetilde{D}_1(T_{obs})$ =0.07   &0.06     &0.58     &0.49  \\
$\widetilde{X}_2$ =0.90  $\widetilde{D}_2(T_{obs})$ =0.06   &0.34     &0.94     &0.15  \\
$\widetilde{X}_3$ =0.96  $\widetilde{D}_3(T_{obs})$ =0.08   &0.42     &0.98     &0.07  \\
\botrule
\end{tabular}}{} 
\end{table}

According to the Kuhn-Munkres algorithm, the best assignment is $\widetilde{R}_1$ with $R_2$, $\widetilde{R}_2$ with $R_1$, and $\widetilde{R}_3$ with $R_3$.
The distance between final states of these two trees is thus 0.58 + 0.34 + 0.07 = 0.99. 
For the remainder of this article, let $D_T$ be the distance between two trees as defined in this section.

\subsubsection{Minimisation of the distance}
\label{sec:Minimisation}

The objective now is to determine the parameter set ($X_0$, $\mu$, $\beta$, $p$, and $L$) that minimizes in average the distance between the simulated trees and the observed genome.
For this purpose, a 5-dimensional grid has been constructed, where each point of this grid represents a parameter set. 

\begin{figure}[ht]
\centering
\includegraphics[scale=0.4]{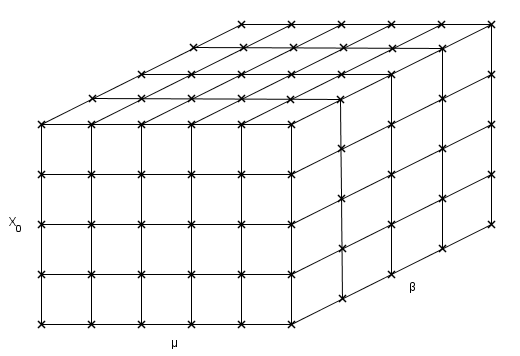}
\caption{A parameter set grid}
\label{fig:Grid}
\end{figure}

A first score $\displaystyle S_1 = \sum_{i = 1}^{N1} D_T(T_i,G)$ has been associated to each of these parameter sets. In this formula, $T_i$ is the $i^{th}$ tree simulated with the parameter set, $G$ represents the observed genome, and $N1$ is a parameter chosen by the user of the optimization method (for instance $N1$ = 100 for the case study in Section~\ref{sec:case studies}).
The best $N2$ points have then been stored and a new score $\displaystyle S_2 = \sum_{i = 1}^{N3} D_T(T_i,G)$ has been associated to each of them. $N2$ and $N3$ are also parameters of the optimization method, with the constraints that $N2$ is lower than the number of points of the grid and that $N3$ is larger than $N1$.
Finally the best of these points are selected to construct a smaller grid around it. This iterative process is continued until the precision chosen by the user of the optimization method has been obtained.

\section{Algorithm}
\label{sec:implementation}

Our proposal has been implemented using Python\footnote{Available at \textbf{https://github.com/SergeMOULIN/retrotransposons-spread}}. A short application programming interface is detailed thereafter.

\subsection{TreeBuild}
This main function is used to build branching trees following the model defined in Section~\ref{sec:general model}. Its halt condition is the targeted number of copies, while its prototype meets the following canvas:  $$(S, T, time) =  TreeBuild (X, mu, B, p, L, n),$$ where $n$ is the desired number of copies, while $X$, $\mu$, $\beta$, $p$, and $L$ are the model parameters as defined in Section~\ref{sec:general model}. $S$ is a $n \times 1$ vector representing states of deterioration. $time$ is the propagation time, also named $T_{obs}$ in this article. Finally $T$ is a $n \times 3$ matrix containing, for each copy: its position, its birth date, and the row of its mother, like in Table~\ref{tab:T}.

\begin{table}[!t]
\processtable{Example of the output $T$\label{tab:T}}
{\begin{tabular}{@{}lllll@{}}
[[  &0.5         & 0.           &0.        &]\cr
 [  &0.19031606  & 1.83699228   &0.        &]\cr
 [  &0.18321005  &11.25706728   &0.        &]\cr
 [  &0.66442132  &17.61532334   &2.        &]\cr
 [  &0.48479738  &25.45993783   &1.        &]\cr
 [  &0.13876928  &28.11662473   &1.        &]]
\end{tabular}}{}
\end{table}

In this example, the mother of the copy located in 0.18 is the root. The mother of the copy located in 0.66 is the copy located in 0.18. Other details regarding this main function are provided thereafter.

\subsubsection{Multiple clocks management}
The working principle of TreeBuild can be summarized as follows: it determines the next event (mutation or duplication) and executes it until the stopping criterion is satisfied. To determine the next event means to know its nature (mutation or duplication), its time, and in which of the available copies it happens. Let $j$ be the number of available copies at time $t_1$. The easiest way to determine the next event is to simulate $2 \times j$ exponential laws, one for each possible mutation or duplication. The minimum of these $2 \times j$ simulations can thus provide the time, the nature, and the copy related to the next event.

Actually, TreeBuild does not really simulate $2 \times j$ exponential laws, as two properties of this law have been used to shorten computations. Indeed, $\forall (p_1, ..., p_{2j}) \in \mathbb{R}^{2j},\forall (Y_1,..., Y_{2j}) \sim \left(\mathcal E(p_{1}), ..., \mathcal E(p_{2j})\right)$, we have:
\begin{enumerate}
\item $\displaystyle min \left(Y_1, ..., Y_{2j} \right) \sim \mathcal E\left(\sum_{i = 1}^{2j} p_i\right)$ ,
\item $\displaystyle \forall i \in 1...2j, P\left(Y_i = min \left(Y_1, ..., Y_{2j} \right)\right) = \frac{p_i}{\sum_{k = 1}^{2j} p_k}$ .
\end{enumerate}
Hence, due to the first property, the time of the next event can be simulated by a single exponential law. The second property, for its part, allows to determine the nature and the copy affected by the next event using a single uniform law.

\subsubsection{Stopping criterion}
\label{sec:Stop condition}
As stated before, the stopping criterion of TreeBuild is related to $n$ (the number of copies of the observed genome). 
But when a genome is observed, there is no way to detect that a new duplication has just occurred.
Thus, the program cannot stop exactly at the birth of the $n^{th}$ copy. Actually, TreeBuild must run until the $T_{n+1}$ iteration (the birth date of the $n+1^{th}$ copy), and then the propagation time $T_{obs}$ can be determined by:
$T_{obs} = T_n + \mathcal U \times (T_{n+1} - T_{n})$,
where $U \sim U[0,1]$.

Furthermore, each value taken by $S$ and $T$ between $T_n$ and $T_{n+1}$ is kept in memory.
Thereby, the values of $T$ and $S$ returned by TreeBuild are values of $T$ and $S$ at time $T_{obs}$.

\subsubsection{The management of copy locations}
\label{sec: Copies location management}
Copy positions in the chromosome are in the interval [0,1]. The distance traveled by a TE before insertion is assumed to follow an exponential law, but this latter can send the new copy outside the interval [0,1]. The solution chosen in this case is to launch again the computation of the new copy position. In other terms:

\begin{center}
While $(X_{child} \notin [0,1]):$

\hspace{2cm} $X_{child} = X_{mother} + U \times Y$ 
\end{center}
where $U \sim U\{-1,1\}$ and $Y \sim \mathcal E(\frac{1}{L})$.

\subsubsection{Critical situations}
When TreeBuild is launched for each point of the grid of parameters, some critical situations can happen, which may induce a significant slowdown of the program. In particular, when $\mu$ is small, and $\beta$ is large, the probability that an event will be a duplication rather than a mutation becomes very low. Thus, TreeBuild executes an inordinate number of mutations before reaching the desired number of copies. Furthermore, the state of deterioration of a copy that faces a great number of mutations converges towards 0.75. To solve this issue, we have decided that, when the state of deterioration of a copy is in [0.74 , 0.76], then this copy cannot mutate anymore.

\subsection{Estimation method}
In the available package, the estimation of the branching model parameters is realized by the \textit{Optim} function. Its prototype is as follows: 
$$Best = optim(Grid, Case, n1, N1, N2, N3).$$ 
Here, $Grid$ is a $5 \times 4$ matrix of settings defined exactly as in Section~\ref{tab:settings}. $Case$, for its part, is a $2 \times n$ matrix containing locations and state of deterioration for each copy of the observed genome. $N1$, $N2$, and $N3$ are settings defined in Section~\ref{sec:Minimisation} and $n1$ indicates how the grid is shrunk at each step after obtaining the best point (\textit{cf.} the following section). The output $Best$ is the parameter set ($ X_0, \mu,  \beta, p,  L$) returned by the \textit{Optim} function.

\subsubsection{Interval reduction}
\label{sec:Intervals reduction}
As explained in Section~\ref{sec:Minimisation}, the estimation method works with a grid where each point represents a parameter set. When the best point of the grid is found, a new grid is constructed around this point. Note that the new grid is not necessarily included in the previous one, in order to provide a larger degree of freedom of the parameters (in particular, when the latter are close to zero). For instance, in the case of parameter $L$, the minimum of the new interval is $min\left(\frac{L_{min}}{2},L_{best} - \frac{L_{delta}}{2 \times n1}\right)$,
where $L_{min}$ and $L_{max}$ are the minimum and maximum of the previous interval, $L_{delta} = L_{max} - L_{min}$, $L_{best}$ is the $L$ coordinate of the best parameter set, and $n1$ is the reduction parameter selected by the user. Thus, the minimum value of the test interval is divided by 2 at each time the best point of the grid is close enough to zero. The maximum value of the new interval is, more simply, $L_{best} - \frac{L_{delta}}{2 \times n1}$. These formulas, written for $L$, are also valid for $\beta$, $\mu$, and $p$. 

\subsubsection{Location in the genome}
Formulas written for $L$ also work for $X$, except that $X$ cannot get out of the interval created by the lowest and highest positions of copies present in the observed genome.
In addition, if $X$ test interval no longer contains any copy position corresponding to those of the observed genome, the nearest from $X_{best}$ is sought. 
Then $X_{best}$ definitely takes the value of this position and is no longer estimated. The estimate continues on the other four parameters only.

\subsection{Module and package dependencies}
Hungarian method has been applied using the ``munkres'' module, implemented in 2008 by Brian M. Clapper \citep{munkres_python}. Furthermore, kernel density estimations used in Part~\ref{sec:results} to estimate $T_{obs}$ have been fitted with \textit{R} software: the function ``density'' of the package ``stat''\footnote{Kernel Density Estimation in R, \textbf{https://stat.ethz.ch/R-manual/R-devel/library/stats/html/density.html}} has been used with all default settings, except the number of equally spaced points where the density is estimated (n = 4096 instead of 512).

\section{Results and Discussion}
\label{sec:case studies}

\subsection{The data}
\label{data}
This proposal has been applied to the spread of the LTR retrotransposons ROO,  DM412, and GYPSY on the chromosome 3L of the \textit{Drosophila melanogaster} genome. This sequence corresponds to the left arm of the chromosome 3, which is the largest autosomal chromosome of \textit{D.~melanogaster}. This is also the most prolific chromosome for each of the LTR retrotransposons we considered, this is why it has been chosen for this case study. 
ROO corresponds to the LTR retrotransposon in \textit{Drosophila melanogaster} with the largest number of copies \citep{kaminker2002transposable, lerat2003sequence, de2009evolutionary}.
DM412 is supposed to have been recently acquired by the \textit{D. melanogaster} through horizontal transfer from a close relative species \citep{bartolome2009widespread, lerat2011comparative, modolo2014new}. 
Finally, GYPSY is an older and likely well regulated LTR retrotransposon \citep{lerat2011comparative}. 
Chromosome 3L contains 32 copies of ROO (with a mean nucleotidic identity of 96\%), 16 copies of DM412 (mean nucleotidic identity of 88.29\%), and 6 of GYPSY (mean nucleotidic identity of 63.84\%). 

Two databases have been used during the experiments. The first one contains positions and nucleotidic sequences for each TE copy annotated in the \textit{D. melanogaster} (flybase website\footnote{\textbf{http://flybase.org/}} version number 5 of the \textit{D. melanogaster} genome \citep{adams2000genome, smith2007release}).
The second database has been downloaded from the RepBase website\footnote{\textbf{http://www.girinst.org/repbase/}} and contains the consensus sequences for each TE corresponding to reference active elements. 
These reference sequences have been placed at the root of the TE tree, corresponding to the sequence state at time 0. Indeed, in the case of ROO, the chromosome 3L contains two copies 100\% identical to the reference sequence, which cannot arrive by chance. This presence justifies to put such a sequence at the root of the tree (however, this is not the case for the two other TEs, that is why an ancestral reconstruction stage should be added to our algorithm, in order to set the root). 

Then the Needleman-Wunsch distance between each TE copy (from the first database) and its reference (from the second database) has been calculated, in order to obtain the deterioration states.  

In this case study, the estimation method described in Section~\ref{sec:Minimisation} has actually been applied not only once but 60 times in each situation, in order to check the consistency of the obtained parameter sets. Then, in each situation, the best set has been determined by the minimization of $\displaystyle S = \sum_{i = 1}^{20,000} D_T(T_i,G)$. The best parameter set of each case study is presented in Section~\ref{sec:results}. The whole obtained parameter sets are presented in supplementary data with their descriptive statistics. Some indications about consistency of these results are provided in Section~\ref{sec:consistency}.

\subsection{Settings}
\label{sec:settings}

Let us first recall that $X_0$, which represents the root position in the chromosome, is inside the interval [0 , 1]. In other words, copy positions have been divided by the chromosome size. 
For the chromosome 3L, this size has been set at 32,600,000 base pairs (bp) according to the sum of euchromatic and heterochromatic region lengths \citep{adams2000genome}.

In Table~\ref{tab:settings}, each row represents the beginning and the end of the test interval, the number by which the test interval has been divided, and the final desired accuracy regarding the parameter.
In particular, in the third column, the value associated to $X_0$, $\beta$, and $p$ is 3. This latter means that these parameters have been tested at the beginning, in the first third, in the second one, and at the end of the test interval. The value associated to $\mu$ and $p$ is 4 (thus these parameters have been tested in 5 points). Finally, at each iteration, the grid contains $4^3 \times 5^2 = 1,600$ points. $\mu$ and $p$ have been evaluated more than the other parameters, as it has been noted in our first analyses that these two parameters have a worse consistency than the other ones.

\begin{table}[!t]
\processtable{Setting table\label{tab:settings}}
{\begin{tabular}{@{}lllll@{}}\toprule \textbf{parameter} &  \textbf{starting point} & \textbf{end point} & \textbf{interval division} & \textbf{desired accuracy}\\ \midrule
   $X_0$   & 0     &1     &3     &$10^{-3}$ \\
   $\mu$   & 0.1   &10    &4     &$10^{-2}$  \\
   $\beta$ & 0.01  &1     &3     &$10^{-3}$ \\
   $p$     & 0.1   &100   &4     &$10^{-1}$ \\
   $L$     & 0.01  &1     &3     &$10^{-3}$ \\ \botrule
\end{tabular}}{}
\end{table}

The other parameters are:
\begin{itemize}
\item $n1$ = 1.5: at each step, after the best point has been found, the grid's dimensions have been divided by 1.5.
\item $N1$ = 100: each point has been tested 100 times.
\item $N2$ = 100: the 100 best points have been re-checked.
\item $N3$ = 1000: the best points have been checked again 1000 times.
\end{itemize}

\subsection{Results}
\label{sec:results}
The obtained parameters are summarized in Table~\ref{tab:results}.

\begin{table}[!t]
\processtable{Results\label{tab:results}}
{\begin{tabular}{@{}llll@{}}\toprule
 \textbf{parameter} & \textbf{ROO} & \textbf{DM412} & \textbf{GYPSY} \\ \midrule
   $X_0$   & 0.024  & 0.652  & 0.728 \\
   $\mu$   & 3.072  & 0.447  & 0.119 \\
   $\beta$ & 0.016  & 0.0023 & 0.028 \\
    $p$    & 0.185  & 0.756  & 0.014\\
    $L$    & 0.435  & 1.516  & $8.400 \times 10^{-4}$\\ \botrule
\end{tabular}}{}
\end{table}

If we consider, for instance, the ROO spread, the obtained parameters can be interpreted as follows:\begin{itemize}

\item $X_0 = 0.0243473400$. Root position in chromosome 3L is $0.0243473400 \times 32,600,000 = 793,723$. The root is the copy located around the $793,723^{th}$ nucleotide.

\item $\mu = 3.072$. The average time between two mutations is 3.072, where 1 is the average time before duplication of the root. Mutations are thus less frequent than duplications. Please note that this estimation of $\mu$ is without time unit: it is related to the duplication speed of the root. It allows to estimate duplication speed when the mutation speed is known, and \textit{vice versa}.

\item $\beta = 0.016$. The proportion of nucleotides affected by a mutation follows a min$(1, \mathcal E(\frac{1}{0.016}))$ distribution. 

\item $p = 0.185$. $p$ allows to determine how many mutations led to a decrease in the duplication speed. For example, in this case, if the identity between a copy and the reference is 0.75 (\textit{i.e.}, state of deterioration = 0.25), then the duplication speed of this copy is reduced by 4.625\% (indeed $0.25 \times 0.185 = 0.04625$).

\item $L$ = 0.435. The distance traveled by the TE before insertion follows a distribution $E(\frac{1}{0.435})$, 
with the constraint that the new copy must remain on the chromosome (\textit{cf.} Section~\ref{sec: Copies location management}).
\end{itemize}

The fact that some of the obtained parameters are outside the test interval chosen at the beginning of the program (for instance, $\beta = 2.32 \times 10^{-3}$ in DM412 case) is a desired effect, to let a larger freedom to the parameters. In particular, the aim was to let parameters to be as close as possible to zero if required (\textit{cf.} Section~\ref{sec:implementation}).

In each of these three cases, one billion of trees have been simulated with the obtained parameter set. The best of these trees is shown in Figure~\ref{fig:ROO_spread} for ROO and in supplementary data for DM412 and GYPSY. Moreover, each simulation returns a propagation time, denoted by $T_{obs}$. Thus, this billion of simulations allowed us to draw histograms or kernel density estimations \citep{epanechnikov1969non} of these propagation times. 
Here, Gaussian kernel density estimations \citep{cleveland1988locally} have been represented on (\textit{cf.} Figure~\ref{fig:Kernel_ROO} for ROO and in supplementary data for DM412 and GYPSY). Propagation times having the highest density are shown in Table~\ref{tab:T_obs}.
As for $\mu$, this estimation of $T_{obs}$ is without time unit, but it is related to the duplication speed of the root.

\begin{figure}[!tpb] 
\centerline{\includegraphics[scale=0.40]{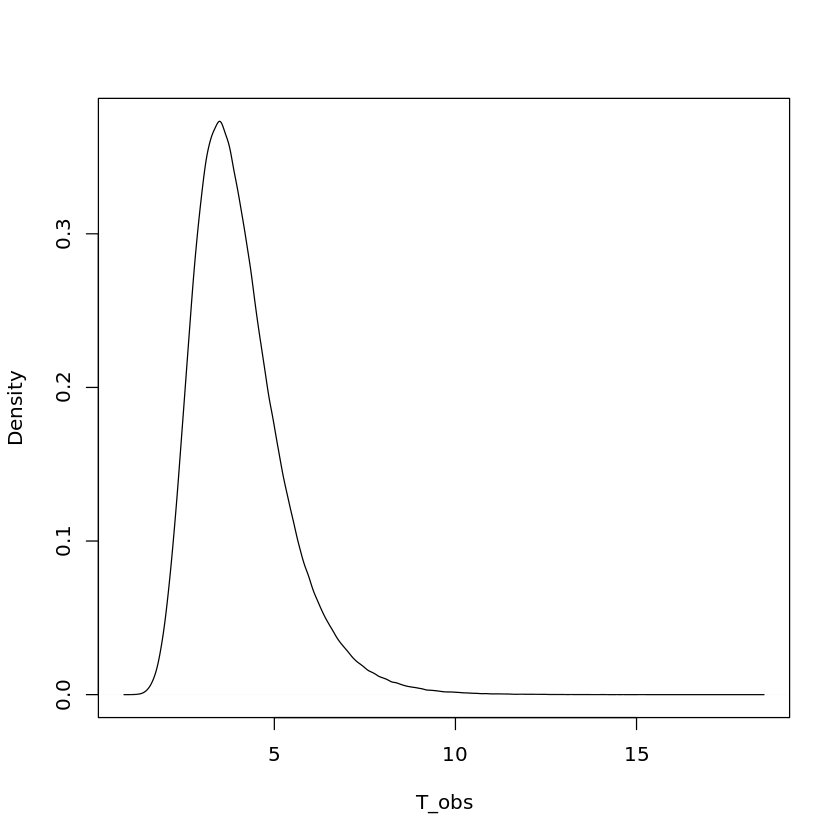}}
\caption{Kernel estimation of $T_{obs}$ for ROO}
\label{fig:Kernel_ROO}
\end{figure}

\begin{table}[!t]
\processtable{Most likely values for $T_{obs}$\label{tab:T_obs}}
{\begin{tabular}{@{}ll@{}} \toprule
 Transposable element & Most likely value for $T_{obs}$ \\ \midrule
 ROO & 3.490  \\ 
 DM412 & 2.705  \\
 GYPSY & 1.671\\ \botrule
\end{tabular}}{}
\end{table}

\subsubsection{Focus on the roots}
According to this model and method, the ROO root could correspond to the FBti0020009 copy. It is an incomplete copy (428 bp, compared to the reference which is 9112 bp length) and corresponds to a solo-LTR, a remnant from a LTR-LTR recombination. This copy is thus no longer active.
The DM412 root could correspond to the FBti0020033 copy. This is a complete copy (7440 bp), which displays 98.8\% identity with the reference. Its two LTRs are identical, indicating that its insertion into the genome is very recent. It is potentially still active since it possess two intact open reading frames encoding the genes gag and pol.
Finally, GYPSY root could correspond to the FBti0060418 copy. It is an incomplete sequence (976 bp compared to the reference which is 7471 bp length) that is very divergent to the reference (78.45\% identity). This copy corresponds to a piece of the inner part of the gypsy element and represents a very old and degraded copy that is not currently active.

\subsection{Consistency of results}
\label{sec:consistency}

As explained previously, the optimization method has been actually applied 60 times for each TE. The descriptive statistics for these three cases are summarized in the supplementary data.
In addition, quotients $\frac{\text{Standard deviation of the results}} {\text{Test interval}}$ have been calculated for each parameter, in each case, in order to assess the consistency of the results. These quotients are reproduced in Table~\ref{tab:Consistency}.

\begin{table}[!t]
\processtable{Consistency indicators\label{tab:Consistency}}
{\begin{tabular}{@{}llllll@{}}\toprule
& $X$ &  $\mu$ & $\beta$ &  $p$ & $L$  \\   \midrule      
 ROO   & 0.092 & 0.223  & 0.013 & 0.162                   & 0.047 \\ 
 DM412 & 0.044 & 0.349  & 0.024 & 0.358                   & 0.180 \\ 
 GYPSY & 0.199 & 0.0047 & 0.012 & $5.290 \times 10^{-4}$  & 0.0016 \\ \botrule
\end{tabular}}{}
\end{table}

\subsection{Investigating other experimental setups}
\label{sec:Others approaches tried}
In this case study, five parameters ($X$, $\mu$, $\beta$, $p$, and $L$) have been estimated in a first step. Then, $T_{obs}$ has been deduced based on its kernel density estimation.
Another approach have been implemented that estimate six parameters together at once. This approach can be summarized as follows: \begin{itemize}
    \item Instead of stopping the tree simulation according to the number of produced copies, the stop condition is $T_{obs} = 1$.
    \item The duplication speed is not fixed at 1 anymore but is a parameter denoted $q$ to estimate. The six parameters to estimate are thus $X$, $\mu$, $\beta$, $q$, $p$, and $L$.
    \item In this approach, the simulated trees will not necessarily have the same number of copies as the observed genome. A penalty for the difference of copy number can be added to the Kuhn-Munkres result.
\end{itemize}

The problem with this approach was that when $q$ is a variable, it is really consistent while any other parameter does not really matter (they are like random). As an illustration, consistency indicators for this approach applied to ROO are presented in Table~\ref{tab:Other approach}, Row 1. As can be seen in this table, the standard deviations of each of these parameters, as a proportion of the associated interval, is very large, while it should be as close as 0 in a consistent scenario.

We have also tried to estimate the whole ($X$, $\mu$, $\beta$, $q$, $p$, and $L$) tuple by minimising: 
$$\displaystyle \sum_{i = 1}^{N} D_T(T_i,G) \mathbf{1}_{\text{number of copies simulated = number of copies in observed genome}},$$ 
where $\mathbf{1}_A$ is the characteristic function of the subset $A$. Consistency indicators of this approach applied to ROO are shown in the second row of Table~\ref{tab:Other approach}. 
For the sake of comparison, these consistency indicators are also given in the case of the approach finally adopted, in the last row of Table~\ref{tab:Other approach}, and with the same computation time. 
Obviously, the latter provide the best results in terms of consistency.
Note that, to achieve this comparison, we have enlarged the precision ($N1$, $N2$, and $N3$ have been increased), which explains why the results contained in the last row of Table~\ref{tab:Other approach} are not the same than what has been previously presented (Table~\ref{tab:Consistency}, Row 1).

\begin{table}[!t]
\processtable{Consistency indicators for the first approaches in the ROO case\label{tab:Other approach}}
{\begin{tabular}{@{}lllllll@{}}\toprule
   &  $X$ &  $\mu$  & $\beta$ & $q$ &  $p$ & $L$  \\   \midrule
 Approach 1     & 0.185 & 0.382 & 0.174 & 0.013 & 0.362 & 0.306  \\
 Approach 2     & 0.112 & 0.408 & 0.255 & 0.014 & 0.473 & 0.300 \\
 Final Approach & 0.079 & 0.204 & 0.013 & NA    & 0.344 & 0.166 \\ \botrule
\end{tabular}}{}
\end{table}

\subsection{Conclusion and future perspectives}
\label{sec:future}
In this article, a model has been proposed for the propagation of LTR retrotransposons in a genome. Various functions have been implemented to simulate this spread as well as graphic representations. Finally, a first method for estimating the parameters of this propagation model has been proposed and applied to the spread of TEs on the ROO, GYPSY, and DM412 elements in a chromosome of \textit{Drosophila melanogaster}.

This work can be improved in various directions, some of them being listed below.

Firstly, it was assumed that an unique root created every copies in the chromosome. The possibility of several roots can be considered too. For instance, a method of unsupervised classification like Gaussian Mixture model could be applied in order to detect the number of clusters.

Secondly, a deletion parameter, which models the disappearance of TE copies, could be considered too.

Furthermore all the parts of the genome are not equally favourable to the TE fixation due their deleterious effects on the neighboring sequences, which result in the elimination of certain insertion by negative selection. It may be appropriate to also take this fact into account.

The effect of an epigenetic regulation that can affect ET behaviour even if they do not face sequence degradation could be taken into account. 

As precised in part~\ref{data}, an ancestral reconstruction stage should be added to our algorithm, in order to set the sequence state of the root at time 0.

Finally, it would be useful to search approaches that allow a faster and more consistent estimation.

\medskip

\textit{Computations have been performed on the supercomputer facilities of the M\'esocentre de calcul de Franche-Comt\'e.
\section*{Funding}
This work has been supported by the Transposable Elements project of the Franche-Comt\'e region.}

\bibliographystyle{natbib}

\bibliography{biblio}
\newpage
\section*{Supplementary data related to the article ``Simulation based estimation of branching models for retrotransposons''}

\begin{figure}[!tpb]
\centerline{\includegraphics[scale=0.45]{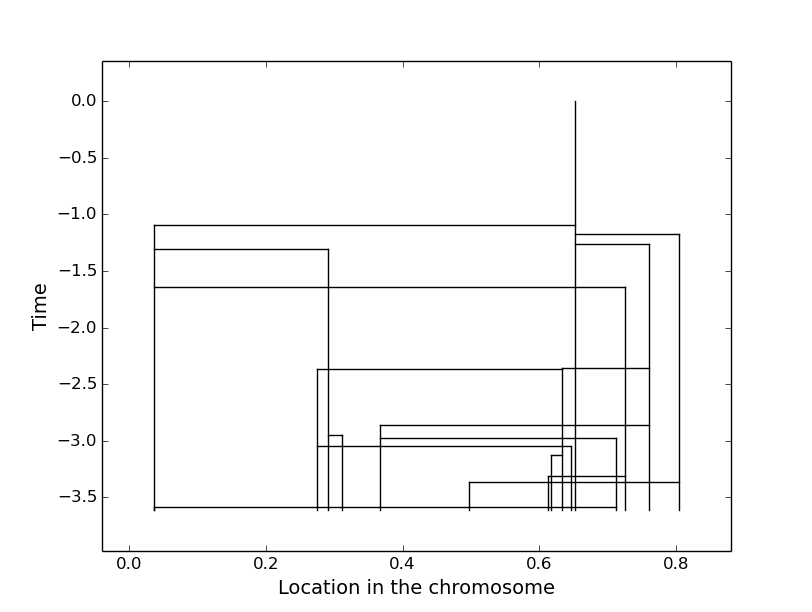}}
\caption{DM412 spread}
\label{fig:DM412_spread}
\end{figure}

\begin{figure}[!tpb]
\centerline{\includegraphics[scale=0.50]{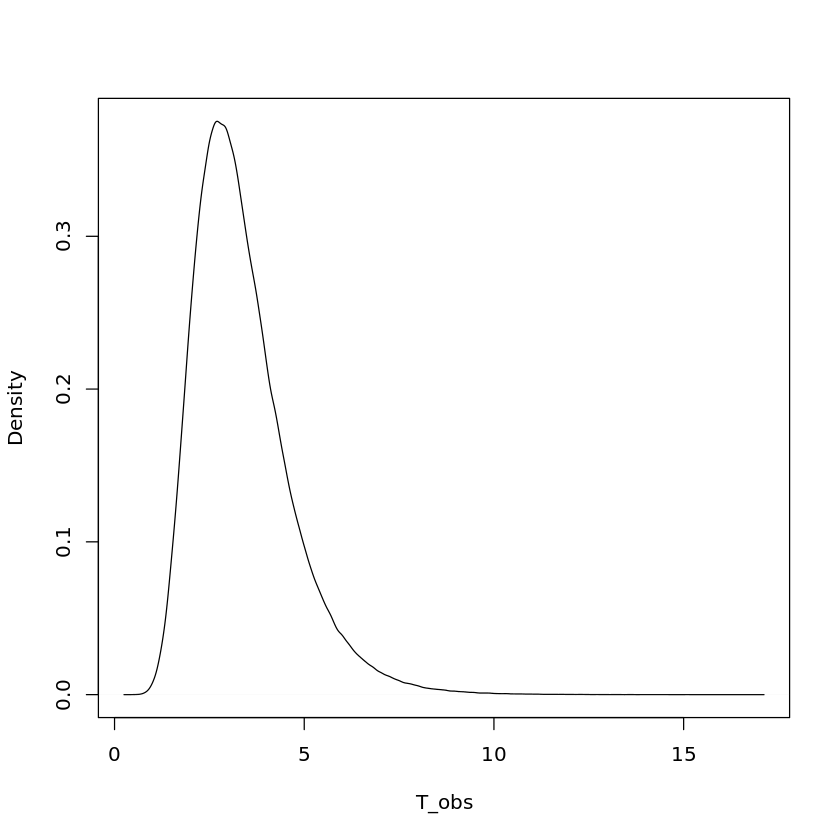}}
\caption{Kernel estimation of $T_{obs}$ for DM412}
\label{fig:Kernel_DM412}
\end{figure}

\begin{figure}[!tpb]
\centerline{\includegraphics[scale=0.50]{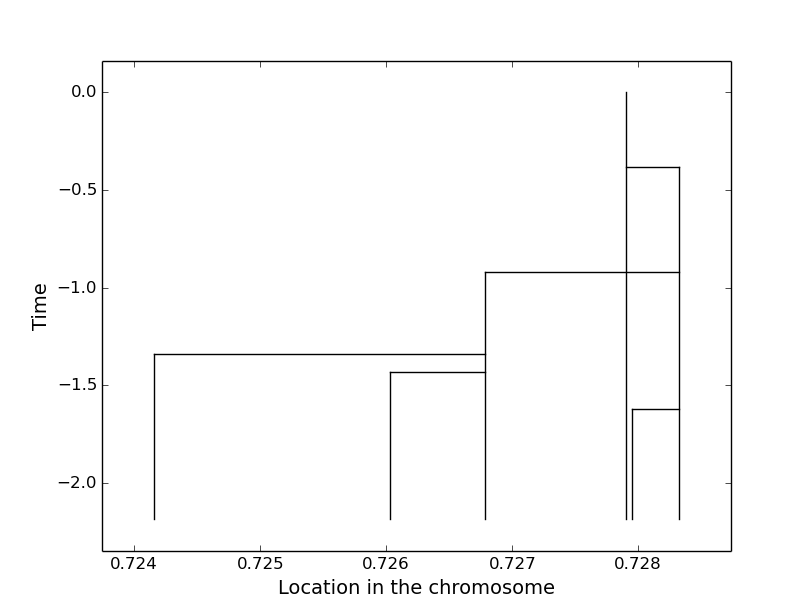}}
\caption{GYPSY spread}
\label{fig:GYPSY_spread}
\end{figure}

\begin{figure}[!tpb]
\centerline{\includegraphics[scale=0.50]{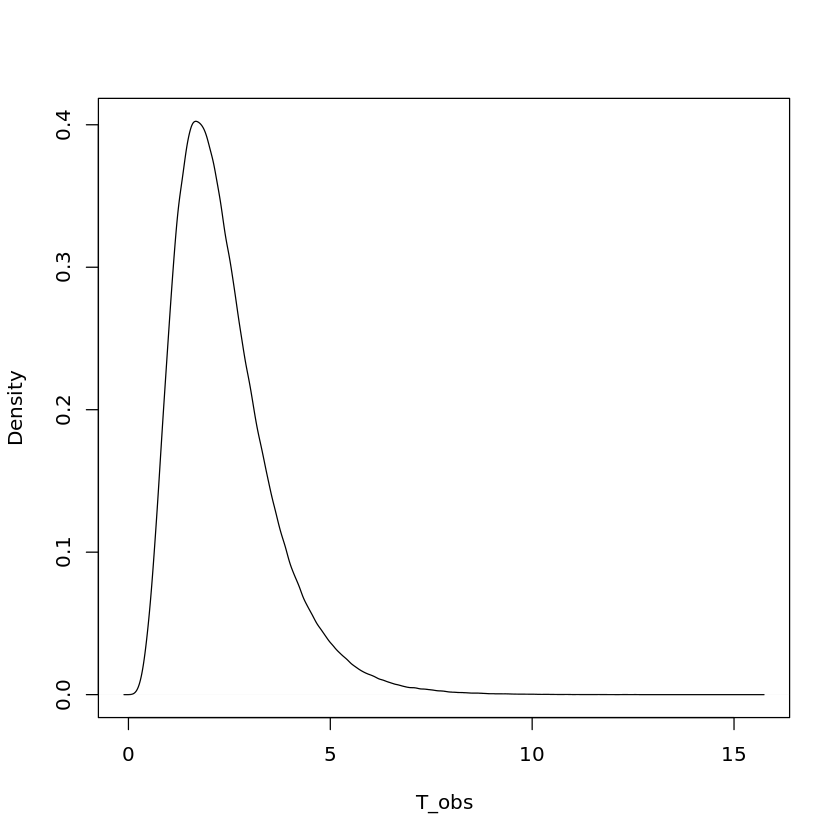}}
\caption{Kernel estimation of $T_{obs}$ for GYPSY}
\label{fig:Kernel_GYPSY}
\end{figure}

This documents provide supplementary data related to the article ``Simulation based estimation of branching models for retrotransposons''.
As explained in the main article, the estimation method has been used 60 times in each situation (\textit{i.e.}, in  ROO, DM412, and GYPSY case).
The descriptive statistics for these 60 trials are summarized in Table~\ref{tab:ROO} for ROO case, Table~\ref{tab:DM412} for DM412 case, and Table~\ref{tab:GYPSY} for GYPSY case.
In addition, the trees that represent DM412 and GYPSY spreads are drawn in Figure~\ref{fig:DM412_spread} and Figure~\ref{fig:GYPSY_spread} respectively. Gaussian kernel density estimations of propagation times for DM412 and GYPSY are finally presented in Figure~\ref{fig:Kernel_DM412} and~\ref{fig:Kernel_GYPSY} respectively.

\begin{table}[!t]
\processtable{ROO, summary of the 60 results\label{tab:ROO}}
{\begin{tabular}{@{}llllll@{}}\toprule
   &   $X$ &  $\mu$ & $\beta$ &  $p$ & $L$  \\    \midrule    
 Min.   &0.0035 &0.332  &$4.941 \times 10^{-4}$    & 0.185  &0.339 \\
 1st Qu.&0.0035 &1.100  &0.0032                    & 6.851  &0.416 \\
 Median &0.030  &2.417  &0.0093                    &16.271  &0.437 \\
 Mean   &0.057  &2.937  &0.013                     &19.963  &0.437 \\
 3rd Qu.&0.075  &3.612  &0.021                     &30.747  &0.461 \\
Max.    &0.211  &8.886  &0.052                     &79.664  &0.586 \\ \botrule
\end{tabular}}{}
\end{table}

\begin{table}[!t]
\processtable{DM412, summary of the 60 results\label{tab:DM412}}
{\begin{tabular}{@{}llllll@{}}\toprule
   &  $X$     &   $\mu$ & $\beta$ &  $p$ & $L$  \\    \midrule      
 Min.   &0.631   & 0.084   &$2.493 \times 10^{-4}$ & 0.7562 & 0.954 \\ 
1st Qu. &0.652   & 0.327   &0.0012                 & 15.432 & 1.091 \\
 Median &0.652   & 0.505   &0.0021                 & 53.094 & 1.250 \\ 
 Mean   &0.666   & 2.334   &0.012                  & 49.176 & 1.238 \\ 
 3rd Qu.&0.652   & 1.732   &0.0088                 & 71.831 & 1.373 \\
 Max.   &0.723   & 11.969  &0.104                  &151.412 & 1.614 \\
\botrule
\end{tabular}}{}
\end{table}

\begin{table}[!t]
\processtable{GYPSY, summary of the 60 results\label{tab:GYPSY}}
{\begin{tabular}{@{}llllll@{}}\toprule
   &     $X$     & $\mu$ & $\beta$      &  $p$ & $L$  \\  \midrule           
 Min    &0.715   &0.074   &0.017   &0.0011 &$2.776 \times 10^{-5}$ \\
 1st Qu.&0.724   &0.123   &0.029   &0.017  &0.0012 \\
 Median &0.724   &0.150   &0.036   &0.031  &0.0022 \\
 Mean   &0.725   &0.155   &0.037   &0.049  &0.0024 \\
 3rd Qu.&0.728   &0.188   &0.044   &0.059  &0.0031 \\
 Max.   &0.730   &0.321   &0.075   &0.305  &0.0069  \\         
\botrule
\end{tabular}}{}
\end{table}

\end{document}